\documentstyle[preprint,12pt]{article}
\def\rz{\mbox{${\sf I \! R}$}}

\begin{document}
\title{Supersymmetric classical mechanics}
\author{Georg Junker$^{\dag}$ and Stephan Matthiesen$^{\ddag}$}
\address{Institut f\"ur Theoretische Physik, Universit\"at
Erlangen-N\"urnberg, Staudtstr.\ 7,\\ 91058 Erlangen, Germany\\[5mm]
$^{\dag}$ e-mail: junker@faupt101.physik.uni-erlangen.de \\
$^{\ddag}$ e-mail: st\_{}matth@faupt101.physik.uni-erlangen.de
}
\shorttitle{Supersymmetric classical mechanics}
\jnl{\JPA\ {\sf to appear}}
\date
\beginabstract
We study the classical properties of a supersymmetric system
which is often used as a model for supersymmetric quantum mechanics. It
is found that the classical dynamics of the bosonic as well as the fermionic
degrees of freedom is fully described by a so--called quasi--classical
solution.
We also comment on the importance of this quasi--classical solution in the
semi--classical treatment of the supersymmetric quantum model.
\endabstract

In 1976 Nicolai [1] introduced supersymmetric quantum mechanics as an
example for the occurrence of supersymmetry (SUSY) in non--relativistic
quantum mechanics.
Independently, in 1981 Witten [2] also suggested SUSY
quantum mechanics as a simplified model for the study of the spontaneous SUSY
breaking mechanism.
The model which has been considered [1-3]
is characterised by the following Lagrangian
\begin{equation}
L:=\textstyle\frac{1}{2}\dot{x}^2-\frac{1}{2}V^2(x)+
\frac{\i}{2}\left(\bar{\psi }\dot{\psi }-\dot{\bar{\psi }}\psi \right)-
V'(x)\bar{\psi }\psi .
\end{equation}
 In the above $x$ denotes a bosonic degree of freedom and hence is an even
Grassmann number. In contrast to this $\psi $ and $\bar{\psi }$ denote
fermionic degrees of freedom and, therefore, are odd Grassmann numbers,
which means that $\{\psi ,\bar{\psi }\}=0$ and $\psi ^2=0=\bar{\psi }^2$.
The real--valued function $V$ is the so--called superpotential.
The Lagrangian (1) describes the supersymmetrized version of a
$(0+1)$--dimensional field theory. In other words, it stems from a
supersymmetric field theory formulated in a superspace spanned by the time
variable $t$ and two Grassmann variables $\varepsilon $ and
$\bar{\varepsilon}$ [1,4,5].
As a consequence, the dynamical system defined by (1)
is invariant under the SUSY transformations:
\begin{equation}
\begin{array}{rcl}
\delta x(t)&=&\varepsilon \psi (t)+\bar{\psi
}(t)\bar{\varepsilon },\\[2mm]
\delta \psi (t)&=&-(\i\dot{x}+V(x))\bar{\varepsilon },\\[2mm]
\delta \bar{\psi}(t)&=&(\i\dot{x}-V(x))\varepsilon.
\end{array}
\end{equation}
The invariance of the system characterised by (1) under the SUSY
transformations (2) is obvious as
\begin{equation}
\delta L=\textstyle\frac{1}{2}\,\frac{\d}{\d t}\left((\dot{x}-\i  V)\varepsilon
\psi +(\dot{x}+\i V)\bar{\psi}\bar{\varepsilon }\right).
\end{equation}
Hence, (2) leads to a gauge--equivalent Lagrangian and, therefore, to
equations of motion being identical to those obtained from the original
Lagrangian (1).

The standard model of SUSY quantum mechanics [2] is found by
quantizing the system (1) either within the canonical
approach [3-5] or the path integral formalism
[3,4]. The increasing interest in this SUSY quantum model has many reasons.
For recent reviews see [6,7]. As a particular motivation, let us mention
the observation that SUSY inspired semi--classical approximations,
the so--called CBC formula [8-11] in the case of
unbroken SUSY and its modification [9-12] for broken SUSY,
yield exact energy eigenvalues for the so--called shape
invariant potentials [13].

In contrast to SUSY quantum mechanics, which has been well studied during the
last ten years, SUSY classical mechanics has, to our knowledge, never been
investigated in details [14].
It is the main purpose of this Letter to present basic results of the
classical system characterised by the Lagrangian (1). In particular,
we will show that the classical solutions for the bosonic as well as the
fermionic degrees of freedom are completely described by the dynamics of a
real--valued {\em quasi--classical} degree of freedom.

The classical equations of motion, which can be derived from the Lagrangian
(1), read
\begin{eqnarray}
\dot{\bar{\psi }}&=&{\rm i} V'(x)\bar{\psi }\\
\dot{\psi}&=&-\i V'(x)\psi\\
\ddot{x}&=&-V(x)V'(x)-V''(x)\bar{\psi }\psi
\end{eqnarray}
where the prime and the dot denote the derivative with respect to $x$ and $t$,
respectively.
The first--order differential equations for the fermionic degrees of freedom
can immediately be integrated. With initial conditions $\psi (0)=:\psi _{0}$
and $\bar{\psi }(0)=:\bar{\psi }_{0}$ the solutions read:
\begin{equation}
\bar{\psi }(t)=\bar{\psi }_{0}\exp\left\{\i\int\limits_{0}^{t}\d\tau
V'(x(\tau ))\right\},\qquad\psi(t)=
\psi_{0}\exp\left\{-\i\int\limits_{0}^{t}\d\tau V'(x(\tau ))\right\},
\end{equation}
where $x(t)$ denotes the (unknown) solution of (6). Let us note that the
solutions (7) imply that $\bar{\psi }(t)\psi (t)=\bar{\psi }_{0}\psi_{0}$
is a constant and, therefore, eq (6) simplifies to
\begin{equation}
\ddot{x}=-V(x)V'(x)-V''(x)\bar{\psi }_{0}\psi _{0}.
\end{equation}
As the superpotential $V$ is assumed to be real--valued the bosonic degree of
freedom $x(t)$, which is an even Grassmann number, necessarily has the
following form:
\begin{equation}
x(t)=:x_{\rm qc}(t)+q(t)\bar{\psi }_{0}\psi_{0},
\end{equation}
where $x_{\rm qc}(t)$ and $q(t)$ are real--valued functions of time. We will
call $x_{\rm qc}(t)$ the {\em quasi--classical} solution in order to
differentiate it from the full classical
solution $x(t)$ which contains the $\bar{\psi }_{0}\psi _{0}$--term.
Only for the
special initial condition $\bar{\psi }_{0}=0=\psi _{0}$ the classical
solution $x(t)$ and quasi--classical solution $x_{\rm qc}(t)$ coincide.
It is also worth mentioning that in the fermionic solutions (7) one may
replace $x(\tau )$ by $x_{\rm qc}(\tau )$ because of (9).

Multiplication of (8) with $\dot{x}$ and integration leads to the energy
conservation
\begin{equation}
{\cal E}=\textstyle\frac{1}{2}\dot{x}^2+\frac{1}{2}V^2(x)+
V'(x)\bar{\psi }_{0}\psi _{0}
\end{equation}
where ${\cal E}$ is a constant even Grassmann number. The ansatz (9)
together with ${\cal E}=:E+F\bar{\psi }_{0}\psi _{0}$ ($E,F\in\rz$) results in
\begin{eqnarray}
&\dot{x}^2_{{\rm qc}}=2E-V^2(x_{\rm qc}),&\\[2mm]
&\displaystyle
\dot{q}=\frac{1}{\dot{x}_{\rm qc}}\left[F-V'(x_{\rm qc})-V(x_{\rm qc})
V'(x_{\rm qc})q\right].&
\end{eqnarray}
The last equation which determines $q(t)$ can also be solved
exactly:
\begin{equation}
q(t)=\frac{\dot{x}_{\rm qc}(t)}{\dot{x}_{\rm qc}(0)}
\left[q(0)+\int\limits_{0}^{t}\d\tau
\frac{F-V'(x_{\rm qc}(\tau ))}{2E-V^2(x_{\rm qc}(\tau ))}\right],
\end{equation}
where $q(0)$ is a constant of integration. Again we find, as for the
fermionic degrees of freedom, that $q(t)$ is expressible in terms of the
quasi--classical solution $x_{\rm qc}(t)$ determined by (11). Let us note that
the singularity of the integral in (13) near the turning points of the
quasi--classical path is precisely canceled by its prefactor as
$\dot{x}_{\rm qc}(t)$ vanishes at those points.
Hence, $q(t)$ remains finite for
all $t\geq 0$. Let us also note that even for the initial condition $q(0)=0$ we
have in general $q(t)\neq 0$ for $t>0$. In other words, even assuming the
classical solution initially to be real, $x(0)\in\rz$, it will in general
become a Grassmann--valued quantity. It is only in the special case $V'(x)=F$,
that is, for a harmonic superpotential, where a real $x(0)$
remains to be real for ever.

Let us now discuss some properties of the quasi--classical solution $x_{\rm
qc}(t)$. The equation of motion (11) for the quasi--classical path can be
obtained from a {\it quasi--classical} Lagrangian defined by
\begin{equation}
\textstyle
L_{\rm qc}:=\frac{1}{2}\dot{x}^2-\frac{1}{2}V^2(x)
=\frac{1}{2}\left(\dot{x}\pm\i V(x)\right)^2\mp\i V(x)\dot{x}.
\end{equation}
The last equality shows that this Lagrangian is gauge--equivalent
to\footnote{The reader should not be confused by the complex gauge
transformation. This defect can be avoided by introducing Euclidean time.}
\begin{equation}
\tilde{L}^{\pm}_{\rm qc}:=\textstyle\frac{1}{2}\left(\dot{x}\pm\i
V(x)\right)^2.
\end{equation}
The canonical momenta obtained from the Lagrangians
$\tilde{L}^{\pm}_{\rm qc}$ are
\begin{equation}
\xi ^{\pm}:=\frac{\partial\tilde{L}^{\pm}_{\rm qc}}{\partial\dot{x}}=
\dot{x}\pm\i V(x)
\end{equation}
and, surprisingly, coincide with the generators of the
SUSY transformation (2) of the fermionic degrees of freedom:
\begin{equation}
\delta \psi (t)=-\i\xi ^-\bar{\varepsilon} ,\qquad\delta\bar{\psi}(t)=\i\xi^+
\varepsilon.
\end{equation}
It is also obvious that the energy $E$ of the quasi--classical solution
can be expressed by $E=\frac{1}{2}\xi ^+\xi ^-$. As a consequence we have the
relation
\begin{equation}
\xi ^{\pm}/\sqrt{2E}=\left(\xi ^{\mp}/\sqrt{2E}\right)^{-1},\qquad E>0.
\end{equation}

As an aside we mention that for $E=0$ the quasi--classical solutions
are given by $x_{\rm qc}(t)=x_{k}$,
where $x_{k}$ are the zeros of the superpotential, $V(x_{k})=0$.
This leads to $\psi (t)=\psi _{0}$, $\bar{\psi} (t)=\bar{\psi} _{0}$ and
$q(t)=0$. Hence, this is the only case where a non--harmonic superpotential
will lead to purely real solutions $x(t)=x_{k}$.

Finally, let us comment on the role played by the quasi--classical path
for the quantum version of model (1). As we have already mentioned, it
has been found that the SUSY inspired semi--classical quantisation does
provide the exact bound--state--energy spectrum for a wide class of
superpotentials. Indeed, it has recently been
verified that these semi--classical quantisation formulas can be derived via
Feynman's path--integral approach [9-11]. The important step in this derivation
was to evaluate the path integral in a stationary phase approximation about
these quasi--classical paths. To be more precise, instead of making the full
action
\begin{equation}
S:=\int\d t L=\int\d t\left[ L_{\rm qc}+
\frac{\i}{2}\left(\bar{\psi }\dot{\psi }-
\dot{\bar{\psi }}\psi \right)-V'(x)\bar{\psi }\psi \right]
\end{equation}
stationary one calculates the corresponding path integral about the
stationary paths of the quasi--classical action
$S_{\rm qc}:=\int\d tL_{\rm qc}$.
This result indicates that the quasi--classical paths (and their quadratic
fluctuations) carry the most important contributions of the path integral.

As a last indication for the dominance of these quasi--classical paths let us
mention an interesting result of Ezawa and Klauder [15]. These authors have
show that as long as one is only interested in expectation values of the
bosonic variable $x(t)$ the path--integral quantisation based on the Lagrangian
(1) is equivalent to that based on (15). To be more explicit, they showed the
relation\footnote{Ezawa and Klauder [15] used Euclidean time.}
\begin{equation}
\begin{array}{l}
\displaystyle\int\left[\prod_{\tau }\d x(\tau )\d\psi (\tau )
\d\bar{\psi }(\tau )\right] x(\tau _{1})\cdots x(\tau _{n})
\exp\left\{(\i/\hbar)\int\limits_{0}^{t}\d \tau  L\right\}\\
\displaystyle\hspace{20mm}=\int\left[\prod_{\tau }\d \xi ^{\pm}(\tau )\right]
x(\tau _{1})\cdots x(\tau _{n})
\exp\left\{(\i/2\hbar)\int\limits_{0}^{t}\d\tau\left(\xi^{\pm}(\tau )\right)^2
\right\}
\end{array}
\end{equation}
where $\xi ^{\pm}(t)$ is defined by (16). It should be stressed that the
last path integral is of Gaussian type. This might be the reason for the
exactness of the SUSY inspired semi--classical approximation. The gauge
transformation $L_{\rm qc}\to\tilde{L}^{\pm}_{\rm qc}$ is the classical
analogue of the Nicolai map discussed by Ezawa and Klauder [15].

\references
\numrefjl{[1]}{Nicolai H 1976}{\JPA}{9}{1497}

\numrefjl{[2]}{Witten E 1981}{\NP}{B188}{513}

\numrefjl{}{\dash 1982}{\NP}{B202}{253}

\numrefjl{[3]}{Salomonson P and Van Holten J W 1982}{\NP}{B196}{509}

\numrefjl{[4]}{Cooper F and Freedman B 1983}{\APNY}{146}{262}

\numrefjl{[5]}{Nicolai H 1991}{Phys.\ Bl\"atter}{47}{387}

\numrefbk{[6]}{Junker G 1994}{Turk.\ J.\ Phys.}{to appear and cond-mat/9403088}

\numrefbk{[7]}{Cooper F, Khare A and Sukhatme U 1994}{preprint}{LA-UR-94-569}

\numrefjl{[8]}{Comtet A, Bandrauk A D and Campbell D K 1985}{\PL}{150B}{159}

\numrefbk{[9]}{Inomata A and Junker G 1993}{Lectures on Path Integration:
Trieste 1991}{ed.\ H A Cerdeira, S Lundqvist, D Mugnai, A Ranfagni, V
Sa--yakanit and L S Schulman (Singapore: World Scientific) p 460}

\numrefbk{[10]}{Inomata A and Junker G 1993}{Proceedings of International
Symposium on Advanced Topics of Quantum Physics}{ed.\ J Q Liang, M L Wang, S N
Qiao and D C Su (Beijing: Science Press) p 61}

\numrefjl{[11]}{Inomata A and Junker G 1994}{\PR}{A}{to appear and
cond-mat/9408054}

\numrefjl{[12]}{Inomata A, Junker G and Suparmi A 1993}{\JPA}{26}{2261}

\numrefjl{[13]}{Gendenshtein L E 1984}{JETP Lett.}{38}{356}

\numrefbk{[14]}{A brief and incomplete discussion has been given in Appendix C
of ref [4]}{}{}

\numrefjl{[15]}{Ezawa H and Klauder J R 1985}{Prog.\ Theor.\ Phys.}{74}{904}

\end{document}